\documentclass[]{iopart}
\usepackage{amstext,amsfonts,amssymb}
\usepackage{mathptmx}  
\usepackage{graphics}

\newfont{\gl}{eufm10 scaled \magstep1} 

\newcommand{\RR}{\ensuremath{\mathbb R}}

\def\ee{{\rm e}}
\begin{document}

\title{\bf{On the number of $k$-cycles in the assignment problem 
for random matrices}}

\author{Jos\'e G. Esteve and Fernando Falceto  } 
\address{ Departamento de F\'{\i}sica Te\'orica, Facultad de Ciencias and \\
Instituto de Biocomputaci\'on y F\'{\i}sica de sistemas complejos.\\
Universidad de Zaragoza,
E-50009 Zaragoza (Spain)
}
\ead{esteve@unizar.es and falceto@unizar.es}
\date{\today}

\begin{abstract}

We continue the study of the assignment problem
for a random cost matrix. We analyse the number of $k$-cycles
for the solution and their dependence on the 
symmetry of the random matrix. We observe that for a
symmetric matrix one and two-cycles are dominant
in the optimal solution. In the antisymmetric case 
the situation is the opposite
and the one and two-cycles are suppressed.
We solve the model for a pure random matrix
(without correlations between its entries) 
and  give analytic arguments to explain
the  numerical results in the symmetric and 
antisymmetric case. We show that the results can be
explained to great accuracy by a simple
ansatz that connects the expected number of
$k$-cycles to that of one and two cycles.

\end{abstract}

\section{Introduction}

The {\it assignment problem} (AP) for a given cost or {\it distance}%
\footnote{We use the term {\it distance} matrix although $d_{{i}{j}}$ are not 
necessarily true distances in a mathematical sense,
in particular they do not need to be positive or symmetric.}
matrix $(d_{{i}{j}}), (i,j=1,\dots, N)$
consists in finding the permutation $\sigma\in S_N$ that minimises
the total distance $\sum_{i=1}^N d_{i\sigma(i)}$. 

There are other problems related to this 
with additional constraints on the permutations allowed.
Probably, the most renowned one is the {\it traveling salesman problem}
(TSP) that can be formulated like the previous AP but
admitting only cyclic permutations 
(we insist that unlike in the standard TSP
our matrix does not need to be a true distance matrix).
The list includes also the {\it minimum weight simple matching 
problem} (SMP)
where only permutations composed of two-cycles are allowed
(obviously in this case $N$ has to be even) and the, somehow opposite case
of  the {\it minimum weight directed  2-restricted 1-factor problem} (1FP), 
for which one-cycles and two-cycles are 
forbidden. If the matrix is symmetric the latter problem can be also 
seen as a {\it minimum weight non directed 2-factor problem} (2FP).
 
{}From the point of view of complexity theory,
it is well known (see \cite{Papa}-\cite{LP}) that the TSP is  NP-hard 
while the 2FP  the AP  and the SMP 
can be solved in a time the scales polynomially with $N$.

In this paper we are interested in the study of the AP for random cost or
{\it distance} matrices.  
 This problem
 has been studied for many years,
 focusing mainly on the minimal 
 distance ${D(AP)}$.
 For example, for random matrices whose entries have probability density 
 $\rho(d_{{i}{j}})= \exp{(-d_{{i}{j}})}\theta(d_{{i}{j}})$ ($\theta$ is the 
Heaviside step function), it was first conjectured by G. Parisi
 \cite{P1} and then proved rigorously (\cite{prueb1}-\cite{prueb3}) that
the expected length is 
\begin{equation}
\langle{D(AP)}\rangle= \sum_{m=1}^{N} {1 \over m^2},
\end{equation}
 with  $N$ the number of points to be matched. Furthermore,
 for general random distances whose densities  behave like  
$\rho(r)=1-a r + {\cal O}(r^2)$ 
near $r=0$, 
it is known (\cite{P21}-\cite{P4}) that
\begin{equation}
\langle{D(AP)}\rangle= \zeta(2)-{2(1-a)\zeta(3)+1 \over N} + {\cal O}(N^{-2}),
\label{DAP}
\end{equation}
where $\zeta(x)$ is the Riemann's zeta function. 

It is also 
known that  for the TSP on symmetric random matrices with $\rho(0)=1$,
 the mean length of the minimal tour is (\cite{atp1},\cite{P5}) 
$D_0 = \lim_{N\to \infty} \langle{D(TSP)}\rangle= 2.041...,$ 
and the next $1/N$ corrections are (\cite{atp2},\cite{atp3})
\begin{equation}
\langle{D(TSP)}\rangle= D_0 \left( 1-{0.1437\over N} -
{10.377 \over N^2} +\cdots\right).
\end{equation}

Different probabilistic relations 
among the problems considered in the previous paragraphs
are also well known in the literature. Namely, since the seminal work of Karp
\cite{Karp} we know that for purely asymmetric random matrices
with uniformly distributed entries we have
$$\lim_{N\to\infty}(\langle{D(TSP)}\rangle-\langle{D(AP)}\rangle)=0.$$
See also \cite{Frieze} and references therein for more precise
estimates of this convergence.

The case of symmetric random matrices is however different, and in this 
situation the expected length of the solution in the TSP and in the 
AP do not coincide in the large $N$ limit. 
A different problem that has been shown to 
be closer to the TSP in probabilistic terms is 
the, above mentioned, 2FP where one-cycles 
and two-cycles are excluded. 
In ref. \cite{Frieze1} it is shown that the expected value 
of the minimal distance for TSP and 2FP with symmetric random matrix
coincides in the large $N$ limit.
These results make clear that the structure of cycles in the optimal 
permutation for the AP depends strongly on the symmetry of 
the distance matrix and 
gives the clue to compare, at a probabilistic level, the different 
related problems.

 Actually, in a recent paper \cite{EstFal}, we found that 
 depending on the characteristics of the distance matrix the AP can interpolate
 between those situations which are near the 
 SM problem (in the sense that the optimal permutation is composed
 approximately of $N/2$ cycles) and those whose 
 optimal permutation is composed of a few cycles 
(just one in some cases) and one and two cycles are absent. 
These can be considered {\it near} the TSP or 2FP solution.
 The transition between both limits is governed by 
 the correlation of the distances $d_{{i}{j}}$ and $d_{{j}{i}}$:
 for positive correlations the AP problem is in
 the ``SM regime'', whereas for anti correlated distances it is
 ``near'' the TSP regime. The transition point is located where there is no
 correlation between the entries $d_{{i}{j}}$
 (that is all the distances are independent random variables), 
a situation that can be solved analytically as we shall see.

In this paper we shall study the expected number of $k$-cycles in the optimal
permutation and its dependence on the symmetry of the distance matrix. 
We shall show  analytic and numerical results with special emphasis in the 
large  $N$ limit. In particular we put into relation the probability of a permutation
to be the solution of the AP with the number of one-cycles and two-cycles 
it contains. This ansatz can account for the numerical results
with high accuracy.

The paper is organised as follows. In the next section we 
describe the problem with full precision. The numerical 
results for the expected value of the number of $k$-cycles
are presented in section 3. In the next three sections
we give analytic arguments to explain the numeric results
in the three regimes: the pure random case, the antisymmetric region 
and the symmetric one. We finally end the paper with some comments and 
conclusions.

\section{Description of the problem}

Given an $N\times N$ matrix $M=(d_{{i}{j}})$ we are interested in the
permutation $\sigma\in S_N$ that minimises the total {\it distance}
$$D_\sigma=\sum_{i=1}^Nd_{i\sigma(i)}$$
This problem is usually named as the {\bf assignment problem} or 
{\bf bipartite matching problem}.
The novelty of our approach is that rather than looking
at the minimum distance itself we focus on the permutation
$\sigma$ that gives this minimum. More concretely we are interested
in the number of $k$-cycles, $p_k$, $k=1,\dots,N$ in the 
permutation $\sigma$ (note that this numbers, determine the 
conjugacy class of $\sigma$ inside $S_N$).

{}From this point of view we shall consider equivalent those
matrices $M$ whose minimum total distance corresponds to  
permutations in the same 
conjugacy class. This implies the following equivalence relation:
\begin{eqnarray}
i)&&\ (d_{{i}{j}})\sim 
(\alpha d_{{i}{j}}+c),\qquad \alpha,c \in \RR,\quad \alpha>0\cr
ii)&&\ (d_{{i}{j}})\sim (d_{\pi(i)\pi(j)})\qquad \pi\in S_N\cr
iii)&&\ M=(d_{{i}{j}})\sim M^t=(d_{{j}{i}}).
\end{eqnarray}

In this paper $M$ is a random matrix that depends on a
constant $\lambda$, we sometimes denote it by $M_\lambda$,
and it is constructed in the following way:
take a random $N\times N$  matrix $R=(R_{{i}{j}})$ whose entries are equally 
distributed, independent, real  random variables with probability density
$\rho$, then the entries of $M_\lambda=(d_{{i}{j}})$ are given by
\begin{equation*}
d_{{i}{j}}=R_{{i}{j}}+\lambda R_{{j}{i}}.
\end{equation*}
Note that, unlike the others, the diagonal 
elements depend on a single random variable
and read $d_{{i}{i}}=(1+\lambda)R_{{i}{i}}$.
Observe that $M_\lambda$ is symmetric for $\lambda=1$,  
antisymmetric for $\lambda=-1$ and purely random (without any correlation
among its entries) for $\lambda=0$.   

{}From the definition of $M_\lambda$ we have
\begin{equation*}
M_{1/\lambda}={1\over\lambda}M_\lambda^t,
\end{equation*}
and,  therefore $M_\lambda\sim M_{1/\lambda}$ for 
$\lambda > 0$ and $M_\lambda\sim -M_{1/\lambda}$ 
for $\lambda < 0$.

As it was mentioned before we are interested in
the number of $k$-cycles $p_k$ or rather
in its expected value in the distribution
generated by $R$, we call it $P_k(\lambda)=\langle p_k\rangle_\lambda$. 
We shall consider $\lambda\in[-1,1]$ that ranges from
the antisymmetric matrix for $\lambda=-1$ 
to the symmetric one for $\lambda=1$.
On the other hand, given the previous equivalence
($M_\lambda\sim M_{1/\lambda}$ for $\lambda>0$),
the results with $\lambda\in(0,1]$ repeat
themselves for $1/\lambda$.
Then in an effective way we cover 
the whole positive real line.
For the negative part things are different
as we have $M_\lambda\sim -M_{1/\lambda}$ for $\lambda<0$;
but, if the probability density for the entries of $R$ is
such that $\rho(x)=\rho(c-x)$ for some constant $c$, then
the distribution of the optimal permutation with $\lambda\in[-1,0)$ 
is again identical to the one for $1/\lambda$.

In the next sections we shall present the results for 
$P_k(\lambda)$ and $\langle n_c\rangle_\lambda$, where $n_c=\sum_k p_k$ is the 
total number of cycles in the optimal permutation.
It is interesting to observe how they change
with $\lambda$ from the antisymmetric point, $\lambda=-1$,
to the symmetric one, $\lambda=1$. 
Different values for the dimension $N$ are considered
to study the large $N$ limit.

We also vary the distribution $\rho$ used to define the model.
We mainly focus on the uniform distribution between $[0,1]$,
with density $\rho_u$, and on the exponential one, 
$\rho_e(x)=\exp(-x)\theta(x)$. 
Note that $\rho_u(x)=\rho_u(1-x)$ and then, in this particular
case, the interval $[-1,1]$ for $\lambda$
is enough to cover the whole
real line. On the other hand, as mentioned in the previous section, 
$\rho_e$ has been extensively used in
studies of the assignment problem for random matrices 
\cite{P1},\cite{P5} which motivates our choice.

The two distributions considered in the previous paragraph
have the same limit for the density in the minimum
of its support $\rho_u(0)=\rho_e(0)=1$. 
Many of the results obtained in the next sections hold independently 
of the distribution used to generate the random matrix provided 
its density function have a non 
zero limit in the minimum of its support. 
The same property is invoked in \cite{P1},\cite{P5}
to have a minimal distance with finite limit when $N$
goes to infinity.    

\section{Numerical results.}

We carried out a numerical simulation of the
statistical ensemble described in the previous section.
For that we generated between $10^5$ and $10^6$
random instances for $M_\lambda$, using the corresponding
probability distributions for the elements $R_{{i}{j}}$. 
The number of instances depends
on the dimension of the matrix, which ranges from $N=40$ to 
$N=1200$.

\begin{figure}[h!]
\resizebox{130mm}{80mm}{
\includegraphics  {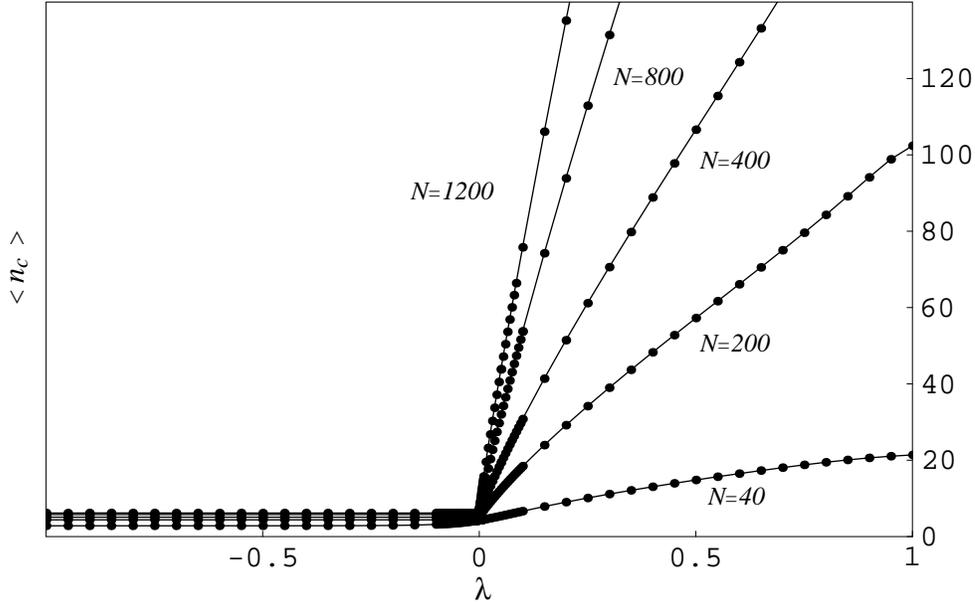}}
\caption{\small Mean value of the number of cycles of the optimal solution for
  the assignment problem at different values of $\lambda$ and $N$.
The dots and the joined plots are obtained with the distributions 
$\rho = \rho_u$, 
 and
$\rho = \rho_e$ respectively. }
\label{fig1}
\end{figure}

Once we generate the matrix $M_\lambda$ we solve the assignment
problem for it using the algorithm of R. Jonker and A. Volgenant
\cite{JV} and compute the number of $k$-cycles $p_k$ obtained in this way.
 In Fig. \ref{fig1} we plot the value of 
$\langle n_c\rangle = \sum_k P_k$; 
there one can see the phase transition between the two regimes of 
$\langle n_c\rangle$ for $\lambda <0$ and $\lambda > 0$. 
In the first case ($\lambda <0$) the expected value of $n_c$
 behaves like $\log (N)$ and is (almost) constant with $\lambda$. 
For $\lambda>0$ the values of $\langle n_c\rangle$ 
grow linearly with $N$ and $\lambda$
\cite{EstFal}.

To understand the behaviour of $\langle n_c\rangle$ in both regimes
we analyse separately the average number of $k$-cycles, $P_k$, 
as a function of $\lambda$ and $k$.
In the rest of the section we present the values
obtained in the numerical simulation.  In the following 
sections we shall give a theoretical explanation of these results.
 
\begin{figure}[h!]
{
\includegraphics{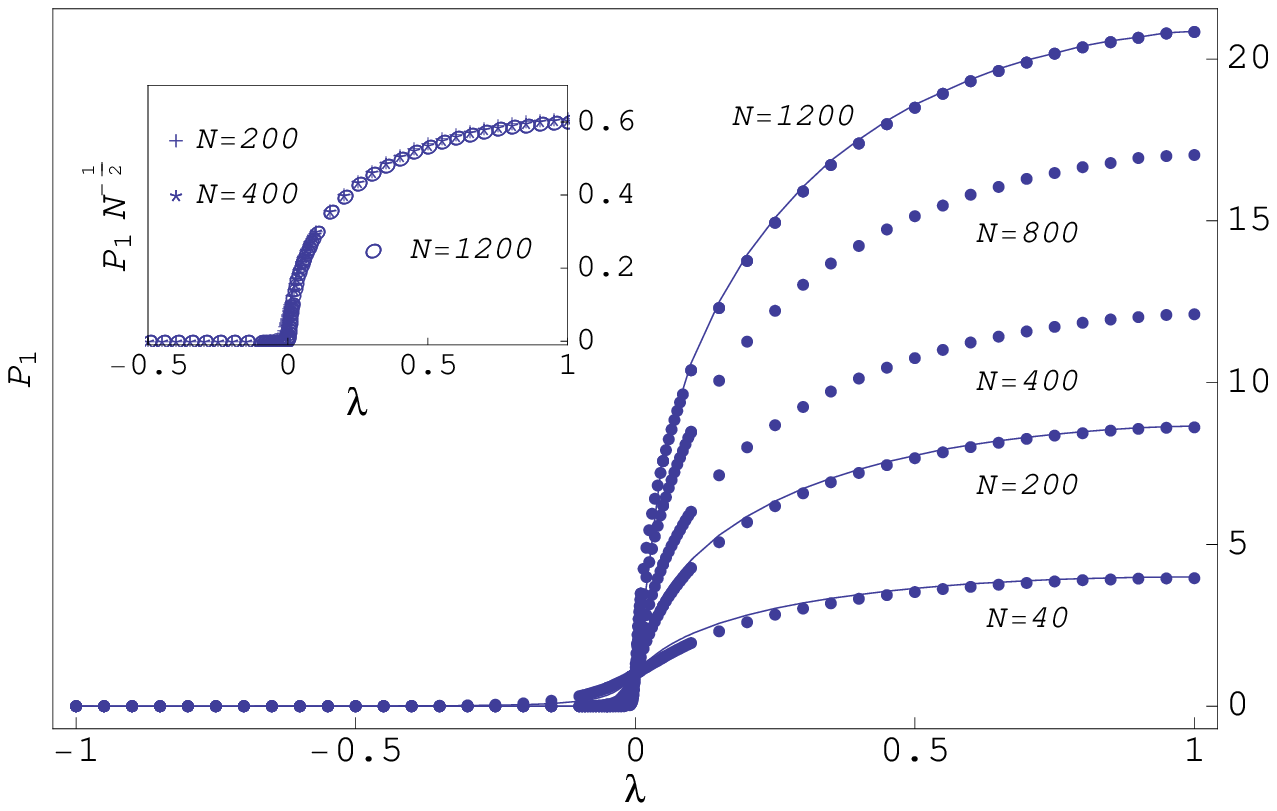}}
\caption{\small Average number of one-cycles in the optimal solution for
  the assignment problem at different values of $\lambda$ and $N$. 
The dots are obtained with the uniform distribution with density 
$\rho = \rho_u$, 
and the joined plots with
$\rho = \rho_e$. Statistical errors corresponding to three standard 
deviation are not visible. The inset shows the behaviour of $P_1 / \sqrt{N}$
 as a function of $\lambda$ for
 different values of $N$.}
\label{p1}
\end{figure}

{\it i) One cycles:}
In the second plot (Fig. \ref{p1}) we show $P_1$ as a function of $\lambda$ 
for different values of the dimension $N$. The dots
correspond to $\rho=\rho_u$ for dimensions $40$, $200$, $400$, 
$800$ and $1200$. The joined plots represent the results for
$\rho=\rho_e$ with $N=40$, $200$ and $1200$. We show no error bars 
because these are negligible. 

We observe that $P_1$ vanishes in all cases in the left part
of the diagram, it attains a common value $P_1=1$ for 
$\lambda=0$ and finally it takes a value 
that grows like $\sqrt N$ for $\lambda = 1$.
We finally note that the joined plots, corresponding 
to a different probability density $\rho=\rho_e$,
lay very close to their respective dots
(for $\rho=\rho_u$) and the fit 
gets better as $N$ grows.

The scaling of $P_1$ with $\sqrt N$ is shown 
 in the inset of Fig. \ref{p1} where  we 
plot $P_1/\sqrt N$  as a function of $\lambda$ 
for different values of $N$. 

\begin{figure}[h!]
\includegraphics {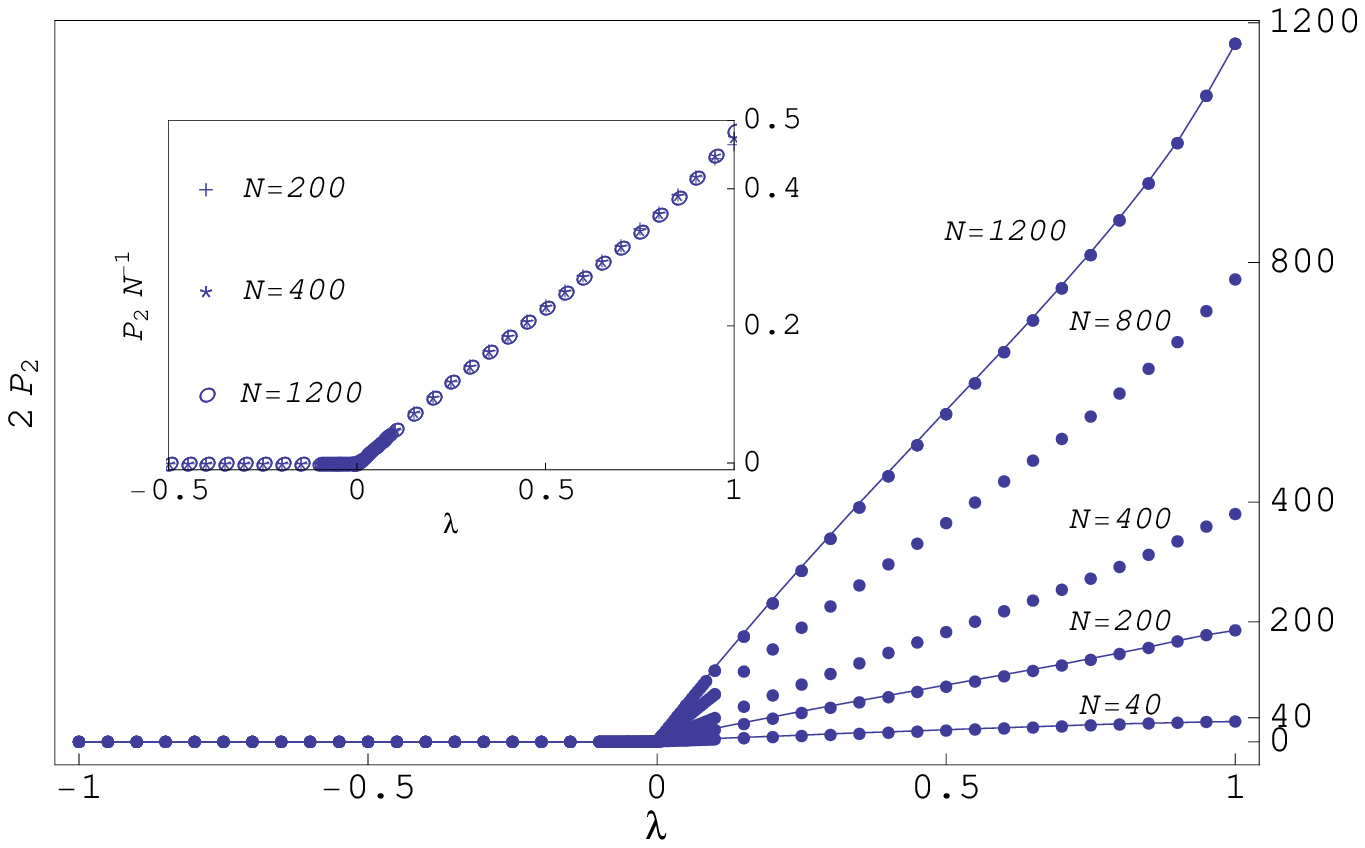}
\caption{\small 
Average number of two-cycles (multiplied by $2$) in the optimal solution for
the assignment problem at different values of $\lambda$ and $N$. 
The results obtained with the densities 
$\rho_u$ ($\rho_e$) are displayed as 
points (joined plot) respectively. Statistical errors are negligible. 
The inset corresponds to 
$P_2 \over N $ versus $\lambda$ for different values of $N$.  }
\label{p2}
\end{figure}

{\it ii) Two cycles:}
In the next plot (Fig. \ref{p2}) we represent $2P_2$ versus $\lambda$.
As in the previous case we show it for different
values of the dimension and different distributions:
the dots correspond to $\rho=\rho_u$ and the joined plots
 to $\rho=\rho_e$. 

We again see that $2P_2$ vanishes near $\lambda=-1$,
takes the value $2P_2=1$ for $\lambda=0$
and grows, in an approximately linear way,
in the symmetric region, $\lambda>0$, to a value
close to $N$ for $\lambda=1$. We also observe that the
points corresponding to $\rho=\rho_u$ fit very well with those
of the joined plot corresponding to $\rho=\rho_e$. 
The inset shows the linear scaling of $P_2$ with $N$, 
for $N\geq 200$.

\begin{figure}[h!]
\resizebox{130mm}{80mm}{
\includegraphics  {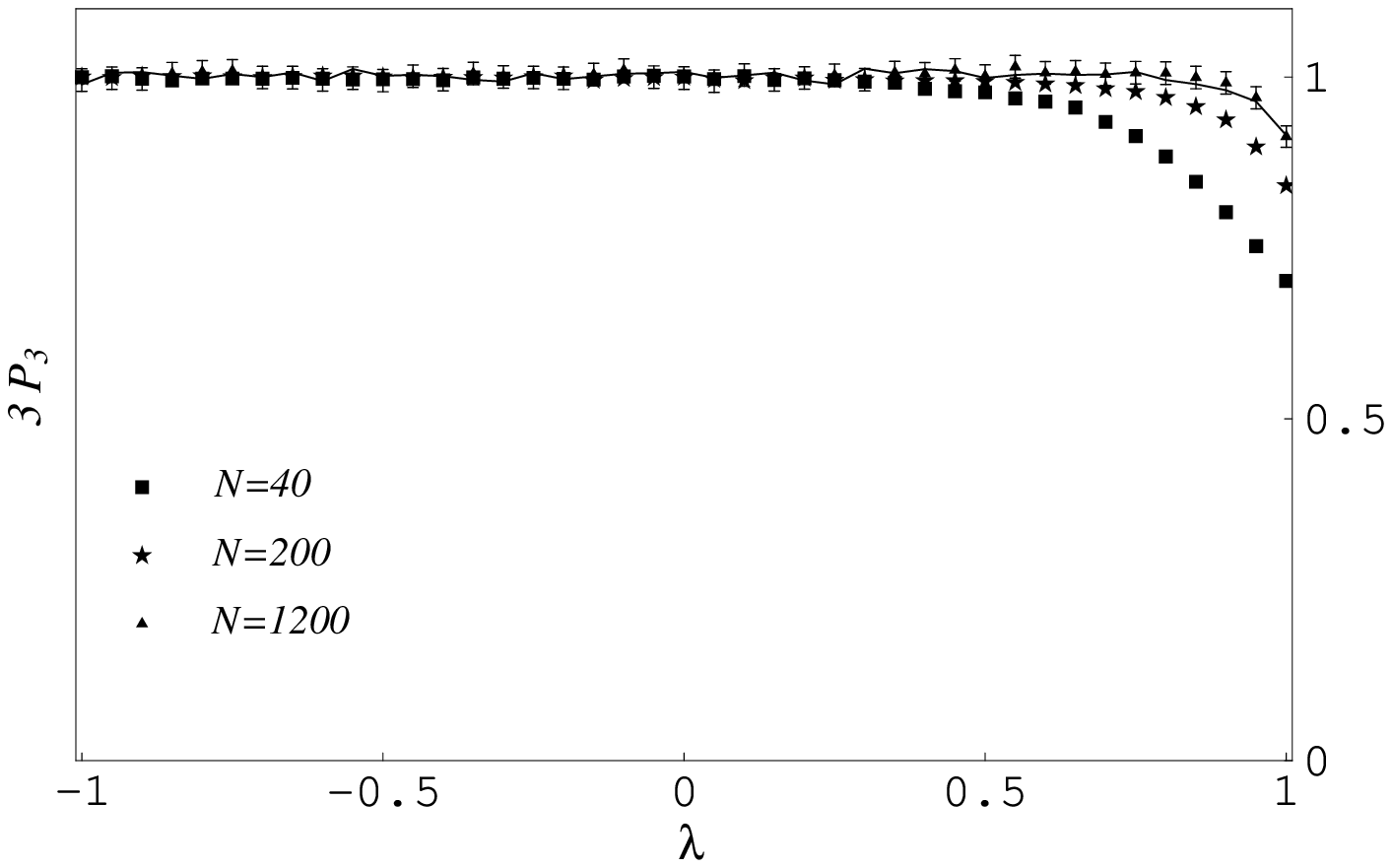}}
\caption{\small Average number of 3-cycles (multiplied by 3) 
in the optimal solution for
the assignment problem.
Symbols correspond to 
$\rho_u$ and different values of $N$ and the joint plot
is for $\rho_e$ and $N=1200$.
The error bars correspond to 
three standard deviations from the mean.
}
\label{p3}
\end{figure}

 {\it iii) Three cycles:}
The situation changes drastically when we plot $3P_3$
as a function of $\lambda$ in
Fig. \ref{p3}.
The dots correspond to $N=40$, $200$
and $1200$ for $\rho=\rho_u$. The joined plot represents the case
of dimension $1200$ with $\rho=\rho_e$. 

We see that $3P_3$ gets a constant value equal to $1$
for almost all values of $\lambda$
and all values of $N$ and $\rho$. Only near $\lambda=1$
things depend on $N$ and as $N$ grows
the value of $3P_3(\lambda=1)$ tends to $1$.  
This limiting behaviour is common for all 
probability densities $\rho$.

Similar results are obtained for other odd cycles
of small length compared to $N$  i.e. $5P_5$ or $7P_7$
are equal to $1$ for all values of $\lambda$ except
near $\lambda=1$, but it tends to $1$ everywhere when $N$
tends to infinity.

\begin{figure}[h!]
\resizebox{130mm}{80mm}{
\includegraphics  {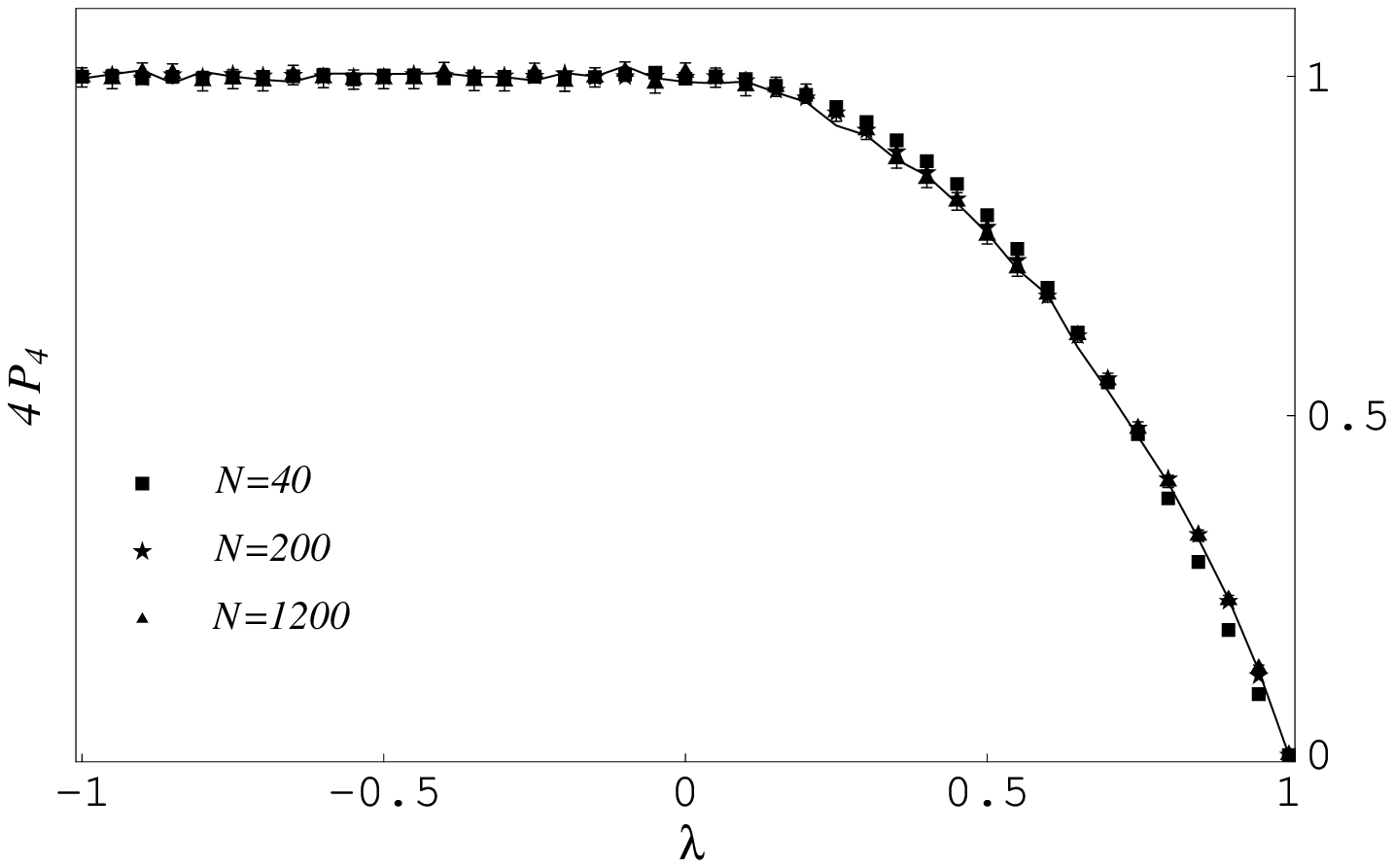}}
\caption{\small Mean value of  4-cycles (multiplied by 4) in the optimal 
solution for the assignment problem.
Symbols correspond to 
$\rho_u$ and different values of $N$ and the joint plot
is for $\rho_e$ and $N=1200$.
Error bars represent three standard deviations.}
\label{p41}
\end{figure}

{\it iv) Four cycles.}
In the next plot (Fig. \ref{p41}) we represent the behaviour of four cycles
plotting $4P_4$ versus $\lambda$ for different values of $N$ and 
$\rho$. Dots represent the values obtained for different
dimensions $N=40$, $200$ and $1200$, 
all with the uniform distribution, with density $\rho_u$. 
The joined plot corresponds to $N=1200$ with $\rho=\rho_e$.

Comparing with the previous plot of $P_3$, we see no change in the left part,
$\lambda < 1$. However the right half is quite different.
We observe that $P_4$ always vanishes at the symmetric point, 
and it follows a smooth curve (even in the large $N$ limit)
from $4P_4=1$ at $\lambda=0$ to $P_4=0$ for $\lambda=1$. 
A similar result is obtained for other {\it short} cycles of even length
like $P_6$, $P_8$, ...:
all of them vanish at $\lambda=1$, only the shape of the curve
changes, it is more horizontal near $\lambda=0$ 
and steeper as we approach the symmetric point.

\begin{figure}[h!]
\resizebox{130mm}{80mm}{
\includegraphics  {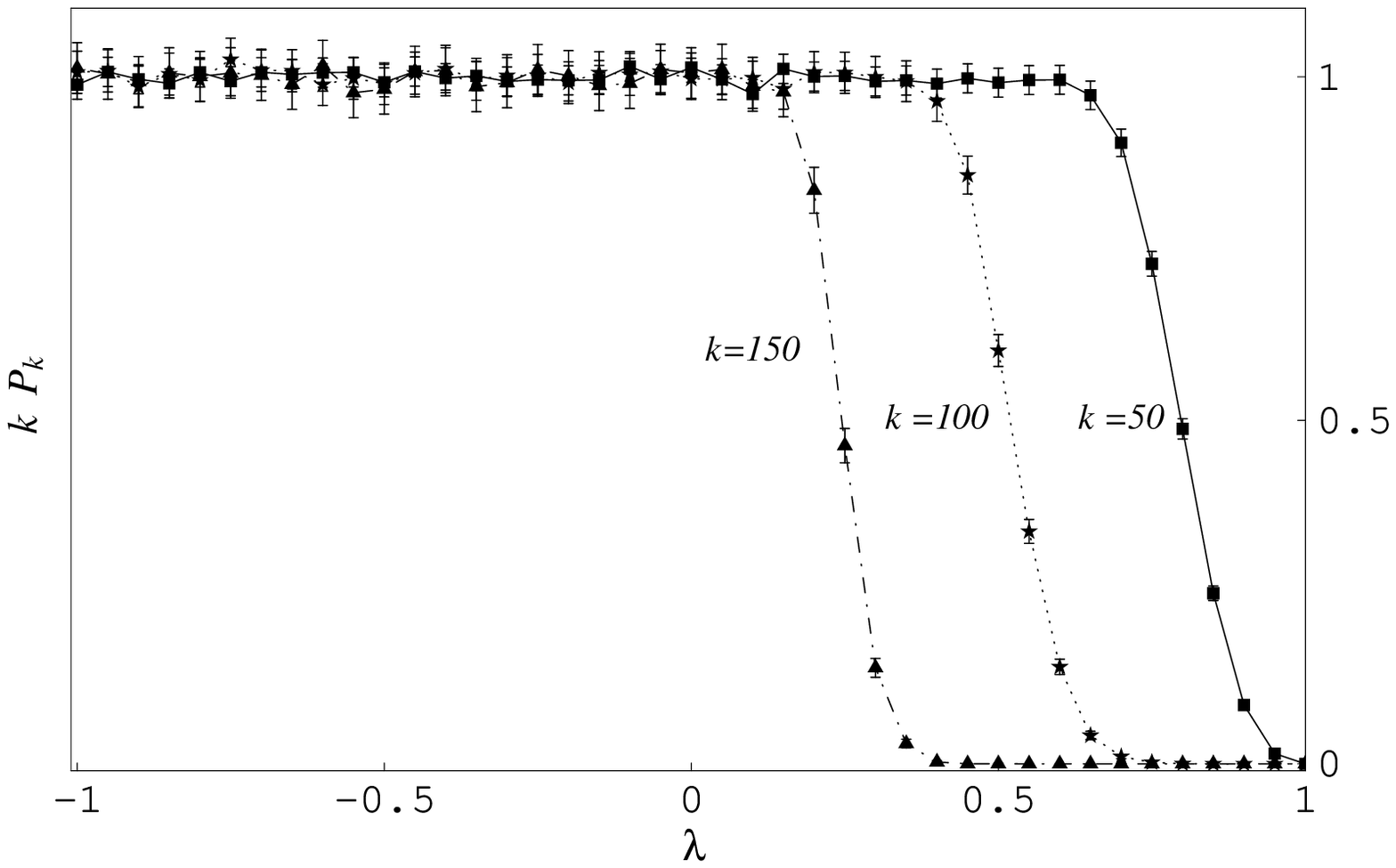}}
\caption{\small Average number of $k$-cycles (multiplied by $k$)
in the optimal solution for the assignment problem at different 
values of $k$ and $\lambda$ for $N=200$. 
Error bars represent three standard deviations. }
\label{pk}
\end{figure}

{\it v) Intermediate cycles:}
In the Fig. \ref{pk} we show the cycles of intermediate length for dimension $200$
and the density $\rho_u$.
As an example we draw $kP_k$ for $k=50,100$ and $ 150$.
We see that, as in previous cases, the behaviour for $\lambda<0$ is 
always constant and equal to $1$. For positive $\lambda$ we see
a fast transition from $1$ to $0$ at a value for $\lambda$ that diminishes
as $k$ increases. Other intermediate values of $k$ and different values of $N$ or $\rho = \rho_e$ 
give similar results (see also Fig. \ref{kpkteo} for odd values of $k$).

\begin{figure}[h!]
\resizebox{130mm}{80mm}{
\includegraphics  {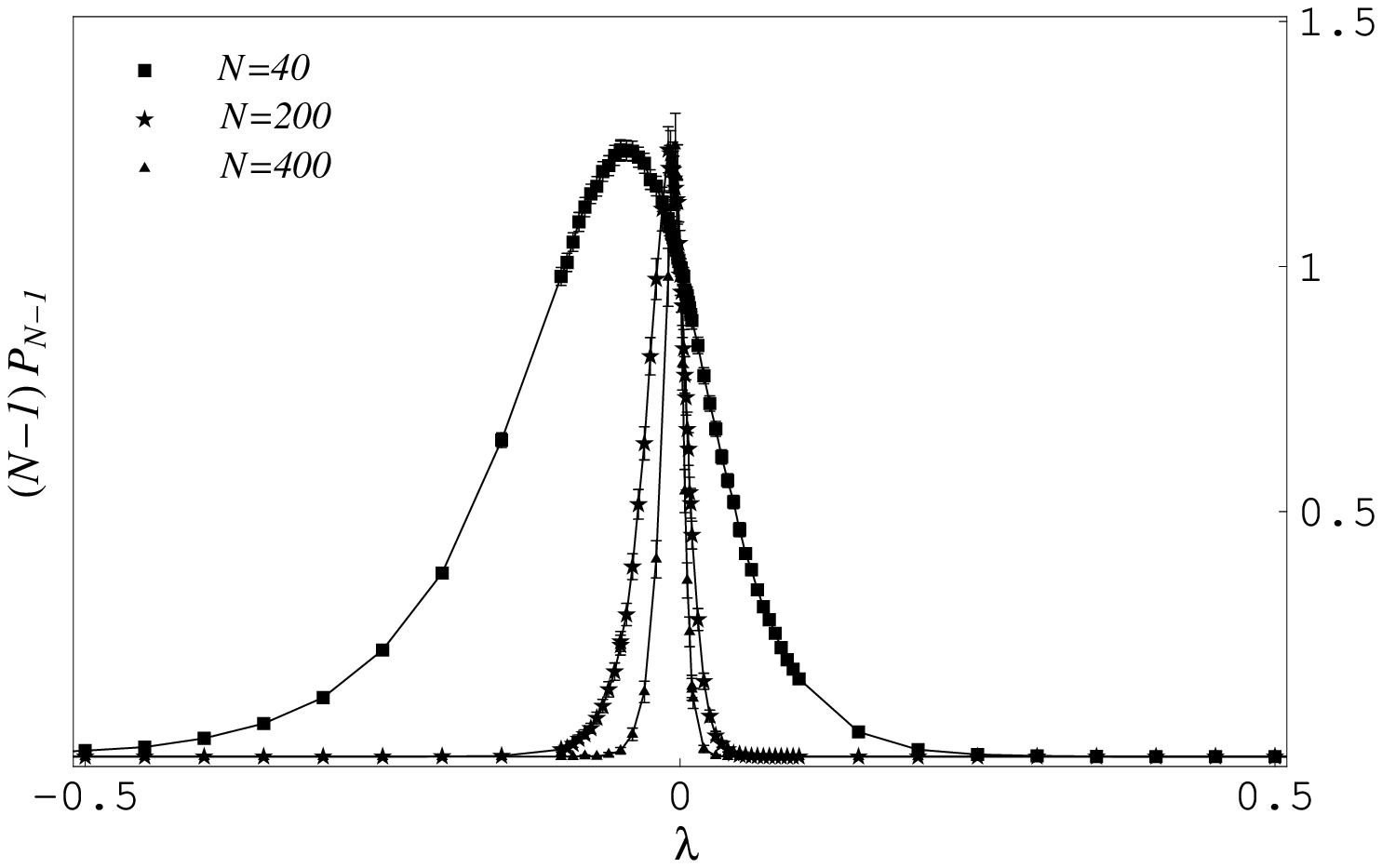}}
\caption{\small Average number of ($N-1$)-cycles (times $N-1$) 
in the optimal solution for
  the assignment problem at different values of $\lambda$ and  $N$.  Error bars represent three standard deviations. Note the common value $(N-1)P_{N-1}=1$ 
 at $\lambda =0$ for all values of $N$.}
\label{pnm1}
\end{figure}

{\it vi) $N-1$ cycles:}
In the Fig. \ref{pnm1} we draw $(N-1)P_{N-1}$ for $N=40, 200, 400$
and $\rho=\rho_u$. We see a peak, sharper as $N$ increases while
its maximum moves toward $\lambda=0$. It always takes
the unit value at $\lambda=0$.
As before, different distributions give similar results.
This plot, as well as those of $(N-2) P_{N-2}$ and $(N-3)P_{N-3}$
which are plotted in the Fig. \ref{teornm2} 
are qualitatively very different from the previous ones
and also different from each other.
In section 5 we shall introduce a simple ansatz that accounts for this,
with great accuracy.

{\it vii) $N$ cycles:}
Finally in the Fig. 8 we present the results for $NP_N$ for different 
dimensions.
Note that it is again constant near $\lambda=-1$
but, contrary to the previous cases, the constant is not $1$ but rather
${\rm e}^{3/2}=4.4816...$. It takes the value 
$1$ for $\lambda=0$ and vanishes for $\lambda>0$. The width 
of the transition is inverse proportional to $N$. Different distributions 
give similar results. 
\begin{figure}[h!]
\resizebox{130mm}{80mm}{
\includegraphics  {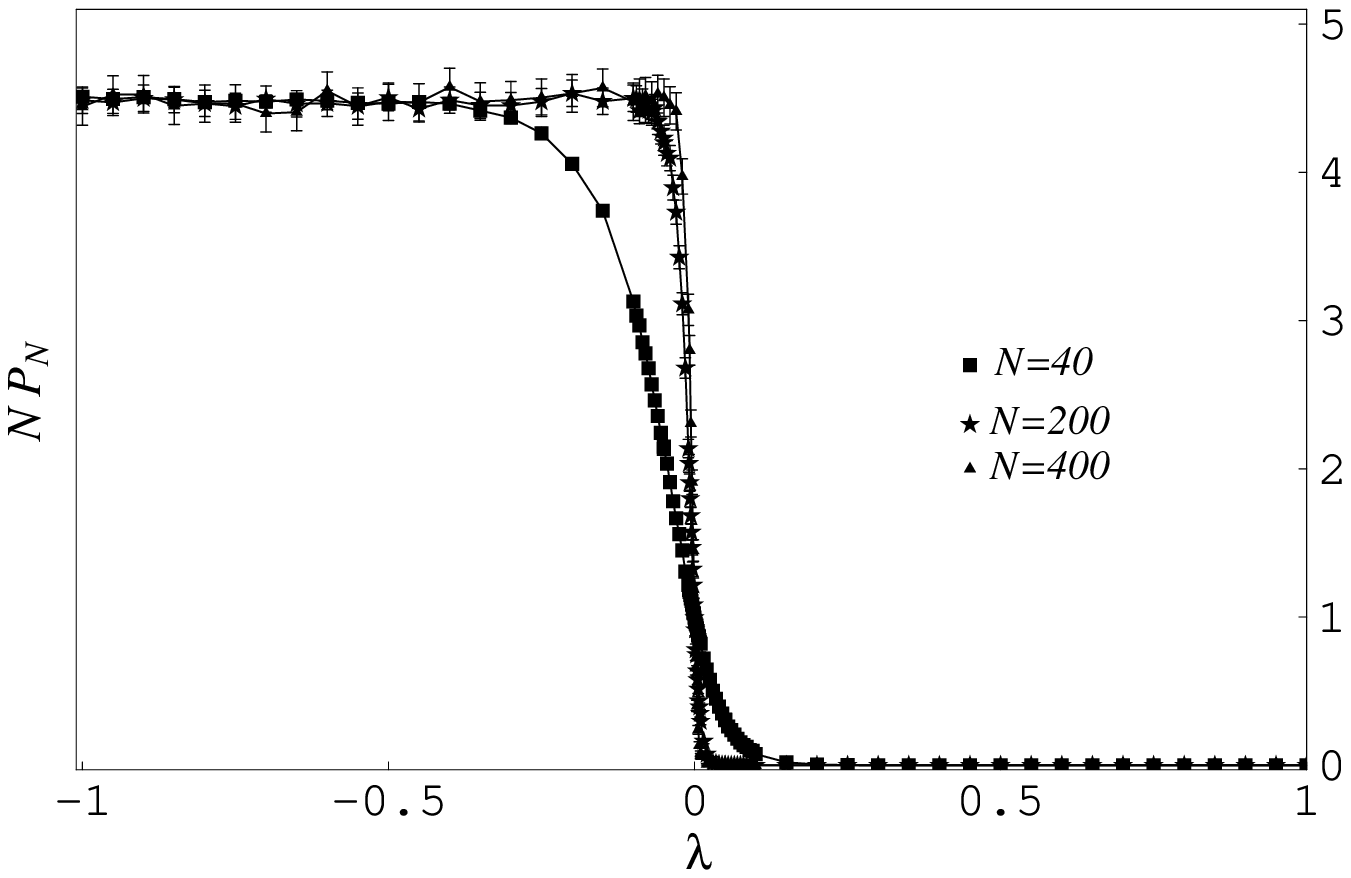}}
\caption{\small Average number of $N$-cycles (times $N$) in the
 optimal solution for
  the assignment problem at different values of $\lambda$ and  $N$.
  Error bars represent three standard deviations.}
\label{pn}
\end{figure}

To summarise the results of this section
we have that for small cycles,
with odd $k>2$,
$kP_k\simeq 1$ for all $\lambda$ in the large $N$ limit.
Small cycles with even $k>2$ have a smooth decay to $0$ at $\lambda=1$. 
For cycles of intermediate length 
$kP_k\simeq 1$ from $\lambda=-1$ until it has an abrupt 
decay at a positive value of $\lambda$
that depends on $k$. 
Cycles of length close to $N$ have a very different behaviour
one from each other. 
And finally, one and two
cycles are absent for $\lambda<0$ and grow
like $\sqrt N$ and $N$ respectively for $\lambda>1$.

\section{Solution of the model for $\lambda=0$.}

We start with the theoretical study of the model
by analysing the point $\lambda=0$. In this case
$M_{0}=R$ and the entries of our matrix
are identical, independent random variables.
Due to this fact we can show that all permutations $\sigma$ 
have the same probability of giving rise to the minimal 
distance. 

The proof is very simple.
Given $M_0=(d_{{i}{j}})$ call 
$${\hat\rho}(M_0)=\prod_{i,j=1}^N\rho(d_{{i}{j}}),$$ 
the probability distribution in the space of matrices
for $\lambda=0$.
It is then clear that
$${\hat\rho}((d_{{i}{j}}))=\hat\rho((d_{\pi(i)j})),$$
for any permutation $\pi\in S_N$.
But if $\sigma$ is the permutation that minimises
the distance $D_\sigma$ for $((d_{{i}{j}}))$ then
$\sigma\circ\pi$ gives the minimum distance for
$(d_{\pi(i)j})$. It implies then that $\sigma$ and $\sigma\circ\pi$
have the same probability of being the optimal permutation,
which leads to the uniform distribution in $S_N$

Once we have established that at $\lambda=0$
all permutations have the same probability
our problem is a purely combinatorial one,
and reduces to compute how many $k$-cycles
there are in $S_N$. This number, that we call $\nu_N(k)$, 
is well known to be 
$$\nu_N(k)={N!\over k}$$
as one can derive from simple counting arguments, i. e.
$\nu_N(k)=\left(\matrix{N\cr k}\right)(k-1)! (N-k)!$ where
the different factors count repectively the possible
choices of $k$ indexes to form the cycle, their orderings
and the permutations of the rest of indexes. Note that in this 
way every permutation is counted as many times as the number of
$k$-cycles it contains, hence the result follows. 
\footnote{We can also use the following iteration 
$\nu_N(k)=(N-k+\delta_{k1})\nu_{N-1}(k)+(k-1)\nu_{N-1}(k-1)$.
The first term in the iteration
counts the number of $k$-cycles that
persist when one add a new index
while the second term stands for  the number of ways
one can add a new index to a $k-1$-cycle to make it 
one unit larger. The $\delta_{k1}$ is there because
for one-cycles, when adding a new index linked to itself 
rather than to any of the preexisting ones, the number of one-cycles 
is increased by one}.

For latter purposes we shall present here
a different, more cumbersome, way to derive $\nu_N(k)$ that makes use of the 
generating function \cite{aSt},\cite{Riordan}. Let
$$G(x)\equiv\sum_{m=1}^\infty {1\over m}x^m = \log\left({1\over 1-x}\right),$$
be the generating function for the number of $k$-cycles in $S_k$
in the sense that
$${d^k\over dx^k}\bigg\vert_{x=0}G(x)= (k-1)!.$$
But we rather want to compute the number of $k$-cycles in $S_N$.
To do this we observe that the generator for the
permutations in $S_N$ are obtained by simply taking the exponential
$$\ee^G=\frac{1}{1-x}.$$ 
The procedure to obtain the number
of $k$-cycles in $S_N$ is then simple. 
We introduce
$$G_\alpha(x)\equiv x+\frac12 x^2+\cdots+
\frac1{k-1}x^{k-1}+
\frac{\alpha}{k}x^k+ 
\frac1{k+1}x^{k+1}+\cdots
$$
so that when we take the exponential of $G_\alpha$
the power of $\alpha$ in every term indicates the number
of $k$-cycles that the corresponding permutation contains.
Therefeore, $\nu_N(k)$ is given by
\begin{eqnarray}
\nu_N(k)&=&\frac{d}{d\alpha}\bigg\vert_{\alpha=1} \frac{d^N}{dx^N}\bigg\vert_{x=0}\ee^{G_\alpha} = \cr
&=&\frac{d^N}{dx^N}\bigg\vert_{x=0}\left(\frac{1}{k}\frac{x^k}{1-x}\right)= 
\frac {N!}{k}.
\end{eqnarray}

The expected number of $k$-cycles for
$\lambda=0$ is then
$$P_k(\lambda=0)=\frac{\nu_N(k)}{N!}=\frac1k.$$
Note that this result is independent of $N$ and of the probability 
density $\rho$ we used to generate the ensemble. 
This explains why in all the results
showed in the previous section $kP_k=1$ for $\lambda=0$. Finally,
 the expected value of
$n_c$ is:
\begin{equation}
\langle n_c\rangle_{\lambda = 0}=\sum_{k=1}^N P_k(\lambda=0)=H_N,
\end{equation}
where $H_N$ is the Harmonic series

\section{The antisymmetric region, $\lambda<0$.}

In this section we study the behaviour
of $P_k(\lambda)$  for $\lambda<0$. We start by the observation
that one-cycles and two-cycles are strongly suppressed
for $\lambda=-1$. The absence of one and two-cycles in the 
solution of the AP makes it equivalent to the corresponding 
1FP as it was mentioned in the introduction. 

This fact can be heuristically understood if one considers that 
the optimal permutation for $M$ comes from the choice 
of $N$ elementary distances $d_{ij}$ out of $N^2$ and, apart from the 
diagonal elements which are $0$, the rest of elements are half of them negative
and half of them positive.
Then, for large $N$, the shortest total distance will be typically
obtained when we chose only negative elements and this excludes
the possibility of having one-cycles ($d_{{i}{i}}=0$)
and two-cycles ($d_{{i}{j}}=-d_{{j}{i}}$) that always include 
non negative entries.
The rest of cycles have no correlation
among their elements and therefore it is
reasonable to assume the equiprobability of all
permutations that do not contain one-cycles or 
two-cycles. With this assumption we reduce the problem to a combinatorial
one and we can proceed like in section 3.

Our goal, however, is to understand the expected number of $k$-cycles
in the whole negative region $\lambda\in[-1,0]$ that interpolates
between the absence of one and two cycles for $\lambda=-1$ 
to the expected values $P_1(0)=1$ and $P_2(0)=1/2$ at $\lambda=0$.
This goal can be achieved with the following ansatz.
We assume that, at least in the large $N$ limit, the probability for a 
permutation to be the 
shortest distance depends only on the number of one-cycles and 
two-cycles it contains.
This is consistent with the fact that only one and two cycles
are sensible to the symmetry of the matrix, bonds of longer
cycles are uncorrelated.
Namely for a permutation with $p_1$ one-cycles and
$p_2$ two-cycles the probability is proportional
to $q_1^{p_1}q_2^{p_2}$, where $q_1$ and $q_2$ vanish
for $\lambda=-1$ and $q_1=q_2=1$ for $\lambda=0$.
The new generating function is then:
$$G_{q_1,q_2}(x)=q_1x+\frac{q_2}{2}x^2+\sum_{k=3}^\infty \frac1k x^k=
\log\left(\frac1{1-x}\right)+(q_1-1)x+(q_2-1)x^2/2.$$
That implements the idea outlined above, as in the exponential 
of $G_{q_1,q_2}(x)$ every term has a weight
$q_1^{p_1}q_2^{p_2}$.
{}From this we derive the normalising factor
(the total weight of the space of permutations)
\begin{eqnarray}
\Omega_{q_1,q_2}(N)
=\frac{d^N}{dx^N}\bigg\vert_{x=0}\ee^{G_{q_1,q_2}}
= \frac{d^N}{dx^N}\bigg\vert_{x=0}\frac{\ee^{(q_1-1)x+(q_2-1)x^2/2}}{1-x},
\end{eqnarray}
while the expected value for the number of $k$-cycles can be 
obtained as in previous section by introducing the factor $\alpha$
multiplying $x^k$ and taking the derivative of the exponential 
at $\alpha=1$.
The result for $k>2$ is
\begin{equation}\label{kcycles}
P_k=\Omega_{q_1,q_2}(N)^{-1}\frac1k
\frac{d^N}{dx^N}\bigg\vert_{x=0}
\frac{x^k\ee^{(q_1-1)x+(q_2-1)x^2/2}}{1-x}\quad\mbox{for}\ {k>2}.
\end{equation}
To compute these quantities we use the singularity analysis
approximation \cite{Flajolet}. In the case at hand the $N^{\rm th}$ 
coefficient in the power series is approximated by the residue at 
the pole in $z=1$. It then gives
$$\Omega_{q_1,q_2}(N)=
N!\left(\ee^{q_1+q_2/2-3/2}+{\cal O}(|q_1-1|^N/N!+
|q_2/2-1/2|^{N/2}/{(N/2)!})\right),$$
and 
\begin{eqnarray}\label{qpk}
&&\hskip -.7cm\Omega_{q_1,q_2}(N)P_k=\frac{N!}k \big(\ \ee^{q_1+q_2/2-3/2}+
{\cal O}(|q_1-1|^{N-k}/(N-k)!+&\cr
&&\hskip 5.7cm+|q_2/2-1/2|^{N/2-k/2}/{(N/2-k/2)!})\ \big).
\end{eqnarray}
For small values of $q_1,q_2$ and $k$ (compared with $N$) this approximation 
can be used and we obtain $P_k\approx\frac1k$ for $k>2$, which is compatible 
with the numerical results of section 3, (see figures 4, 5 and 6). 

$P_1$ and $P_2$ do not follow the general formula but
\begin{eqnarray}\label{eqPuno}
P_1=\Omega_{q_1,q_2}(N)^{-1}
\displaystyle\frac{d^N}{dx^N}\bigg\vert_{x=0}
q_1x\ee^{G_{q_1,q_2}(x)}
\end{eqnarray} 
which, in the singularity analysis approximation, gives
\begin{equation}
P_1=q_1+{\cal O}(|q_1-1|^{N-1}/(N-1)!+|q_2/2-1/2|^{N/2-1/2}/(N/2-1/2)!).
\label{pq1}
\end{equation}
And 
\begin{eqnarray}\label{eqPdos}
P_2=\Omega_{q_1,q_2}(N)^{-1}
{\displaystyle\frac{d^N}{dx^N}\bigg\vert_{x=0}
\frac{q_2}2 x^2\ee^{G_{q_1,q_2}(x)}}
\end{eqnarray} 
so that
\begin{equation}
P_2=\frac{q_2}2+
{\cal O}(|q_1-1|^{(N-2)}/(N-2)!+|q_2/2-1/2|^{N/2-1}/(N/2-1)!).
\label{pq2}
\end{equation}
Then for small values of $q_1$ and $q_2$ and the values of
$N$ we are considering in the paper (from $40$ to $1200$)
we can take $P_1=q_1$ and $P_2=q_2/2$ with a very good 
accuracy (that covers the $\lambda < 0$ region since there $q_1$ and 
$q_2$ are less than $1$).

For long cycles $k\sim N$ the singularity analysis approximation 
is not valid any
more. In this case, however, it is very easy to compute (\ref{kcycles}) 
explicitly.
Therefore with the precision given by that of 
$\Omega_{q_1,q_2}(N)$ we get:
\begin{eqnarray}\label{longcycles}
 \hskip 1.cm N P_N &\simeq&\ee^{3/2-q_1-q_2/2}\cr
(N-1)P_{N-1}&\simeq&q_1\ee^{3/2-q_1-q_2/2}\cr
(N-2)P_{N-2}&\simeq&(q_2/2+q_1^2/2)\ee^{3/2-q_1-q_2/2}\cr
(N-3)P_{N-3}&\simeq&({1/3}+ q_1 q_2 /2 +  q_1^3 /6)\ee^{3/2-q_1-q_2/2}
\end{eqnarray}

\begin{figure}[h!]
\resizebox{130mm}{80mm}{
\includegraphics {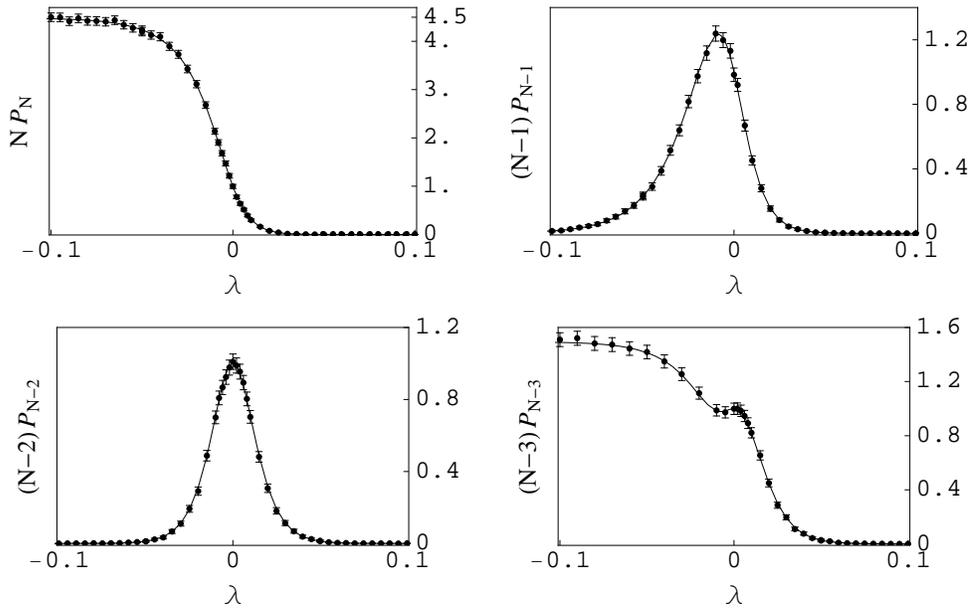}}

\caption{\small Average number of $k\, P_k$ (for the largest values of $k$) 
in the optimal solution for
  the assignment problem at different values of $\lambda$ and for $N=200$.
  The points are the result of our simulation and the error bars represent
 three standard deviations from the mean. The joined plot is the theoretical
 prediction  using  (\ref{longcycles}).}
\label{teornm2}
\end{figure}

In Fig. \ref{teornm2} we plot $kP_{k}$ for $k=N,\cdots , N-3$ and  $N=200$. The continuous
line is the theoretical value obtained from (\ref{longcycles})
where we take $q_1=P_1$ and $q_2=2P_2$. One can see
that the agreement is excellent. A similar match holds for the 
other cases.

Thus, from the previous expressions
we see that the behaviour of $P_k$ for for $k=1,\dots, N$  
for $\lambda\leq0$
is completely determined by $q_1$ and $q_2$. In the rest of the section 
we shall study the behaviour with $\lambda$ and $N$ of this two factors.
Many of the results presented below
are independent on the distribution 
used to generate the random matrices, provided
the probability density fulfils the non vanishing
property in the minimum  of its support that was discussed
in section 2. In the rest of the paper we shall assume that this property
holds.

\begin{figure}[h!]
\resizebox{130mm}{80mm}{
\includegraphics  {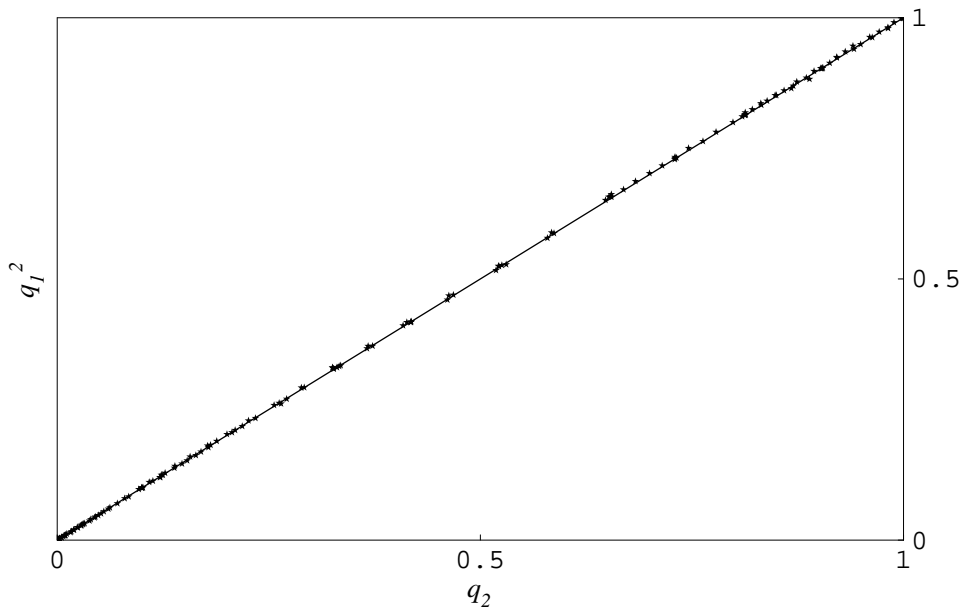}}
\caption{\small Values of $q_1^2$ versus $q_2$ for $N=40, 200, 800$ and
 $\rho=\rho_u$, $\rho_e$. The continuous plot is the line $q_2=q_1^2$.}
\label{q1q2}
\end{figure}

Our first observation is the relation between $q_1$ and 
$q_2$ for the same value of $N, \lambda$ and $\rho$.
One can check that $q_2=q_1^2$.
A plot showing the extremely good fit between the two values as a function of
$\lambda<0$ for $N=200$ and different $\rho$ is shown in Fig. \ref{q1q2}.
This relation can be expressed as the fact that the probability of
a permutation to produce the minimal total distance,
is unchanged if we change the permutation by substituting 
a two-cycle by two one-cycles. 
An argument for this comes from the fact
that given two indexes $i$ and $j$, $d_{{i}{j}}+d_{{j}{i}}=
(1+\lambda)(R_{{i}{j}}+R_{{j}{i}})$ while 
$d_{{i}{i}}+d_{{j}{j}}=(1+\lambda)(R_{{i}{i}}+R_{{j}{j}})$. Then both
sums are identical random variables.

\begin{figure}[h!]
\resizebox{130mm}{80mm}{
\includegraphics  {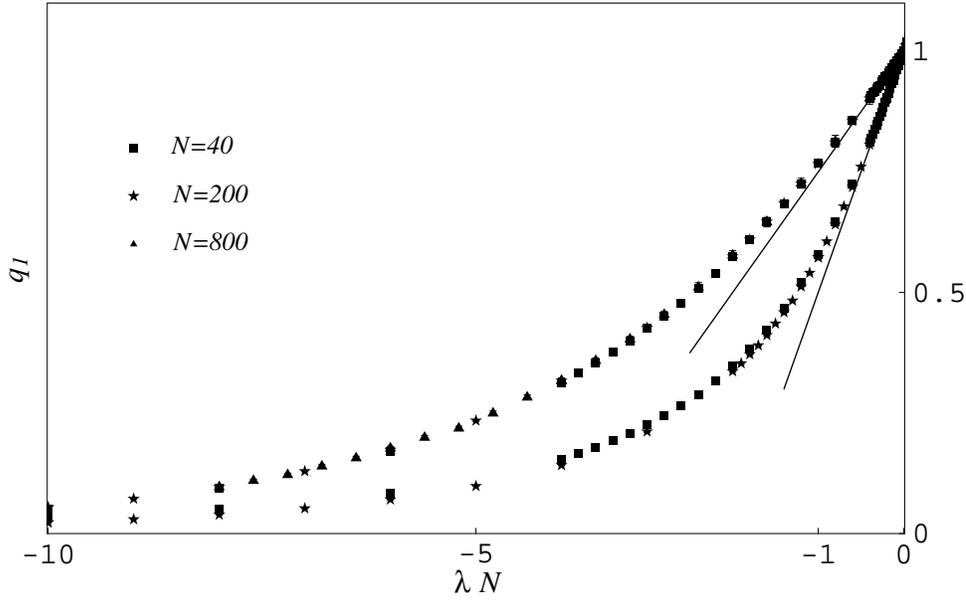}}
\caption{\small The points in the upper curve represent the  values of $q_1$ 
as a function of $\lambda N$ for different values of $\lambda$
 and $N$ and for matrices generated with probability density $\rho_u$. The 
 tangent line at $\lambda =0$ is the  theoretical prediction given by
 (\ref{slope}). The lower curve is the same
 but for matrices generated with the exponential density$\rho_e$.}
\label{q1}
\end{figure}
The second important property we observe in the region $\lambda<0$ 
is the invariance under scaling of $\lambda$ and $N$ (see Fig. \ref{q1}). In fact 
one can check that for a given probability density $\rho$, 
$q_1(\lambda,N)=q_1(\mu\lambda, \mu^{-1}N)$.
And as any $P_k$ can be obtained
from $q_1$ according to the formulae above, this 
scale invariance is true also for any $P_k$.

The scaling relations presented in the previous paragraph are obtained  
by taking a fixed probability density $\rho$
to generate the ensemble, while we change 
$\lambda$ and $N$. We want to examine now how
$q_1$ depends on the distribution near
the random point $\lambda=0$. Given the result that we can 
rescale $\lambda$ and $N$ without changing $q_1$
it is natural to think that $q_1$ can be determined by 
looking at only a few elements of the matrix $M_{\lambda}$.
A confirmation of this conjecture is not available yet, but
some partial results can be verified. Concretely
we can reproduce the slope of $q_1$ at $\lambda=0$,
that depends on the distribution,
by the following formula:  
$$\frac{\partial q_1}{\partial\lambda}(\lambda=0,N)=\alpha N.$$
Where $\alpha$ depends solely on the distribution and is 
determined as follows:\hfill\break
for a given value of lambda fix $i\not=j$ 
and define $\xi_E={\rm min}(d_{{i}{j}},d_{{j}{i}})$, 
also define $\xi_D={\rm min}(d_{{i}{i}},d_{{j}{j}})$. Now 
compute 
$$\Theta(\lambda)\equiv\frac12\langle\theta(\xi_D-\xi_E)\rangle_\lambda,$$
where with $\theta$ we denote the Heaviside step function. The
coefficient $\alpha$ is obtained by
$$\alpha=-\frac{\rm d}{{\rm d}\lambda}\Theta\vert_{\lambda=0}.$$ 
As we mentioned before the value of $\alpha$ depends only on the 
probability density $\rho$ and can be computed with the following formula
\begin{eqnarray}\label{slope}
\alpha=2\int_{-\infty}^\infty \rho^2(x)\int_x^\infty\rho(y)
\int_x^\infty(z-x)\rho(z){\rm d}z {\rm d}y {\rm d}x.
\end{eqnarray}
The meaning of $\Theta$ is the following: it measures the probability
for an extra diagonal element of a pair to be  smaller than its pair
and than two entries in the diagonal.  
It, somehow, reproduces at a small scale (only four random variables 
involved) the mechanism for the disappearance of one-cycles 
(diagonal entries) in the real problem as $\lambda$ starts to be 
negative. 
Recall that the argument for the disappearance of one and two-cycles
was based in the fact that for negative $\lambda$
one of every pair of extra diagonal terms is smaller
(in average) than the diagonal terms
(or than half the sum of the extra diagonals).
It then implies that the appearence of one and two-cycles 
in the optimal permutation is disfavoured. This property is 
quantitatively studied by means of the function $\Theta$.

Our result has been checked with different distributions and
the agreement is very good. As an example we show in Fig. \ref{q1} 
the lines for $\rho_e$ and $\rho_u$ with slope
$1/2$ and $1/4$ respectively, as obtained from (\ref{slope}). 
We can see that these lines are, as predicted,
tangent to the curve of $P_1$ at $\lambda=0$.

\section{The symmetric region $\lambda>0$}
As shown in Fig. \ref{p1} and \ref{p2}, the first relevant 
fact in this region
is that $P_1$ and $P_2$ grow from $1$ and $1/2$ respectively 
for $\lambda=0$, to values proportional to $\sqrt{N}$ 
in the first case and to $N$ in the second for $\lambda=1$.
A first attempt to account for this behaviour
is to adjust the corresponding parameters
$q_1$ and $q_2$ to fulfil equations (\ref{eqPuno}) and 
(\ref{eqPdos}), (note that now $q_1$ and $q_2$ can be $\gg 1$ so the terms
of order $ (q_1-1)^{(N-2)}/(N-2)!$ and $(q_2/2-1/2)^{N/2-1}/(N/2-1)!$ can be important).
 The values of $q_1$ and $q_2$ obtained
in this way are used to compute $P_k$ for 
different values of $\lambda$. 

\begin{figure}[h!]
\resizebox{130mm}{80mm}{
\includegraphics  {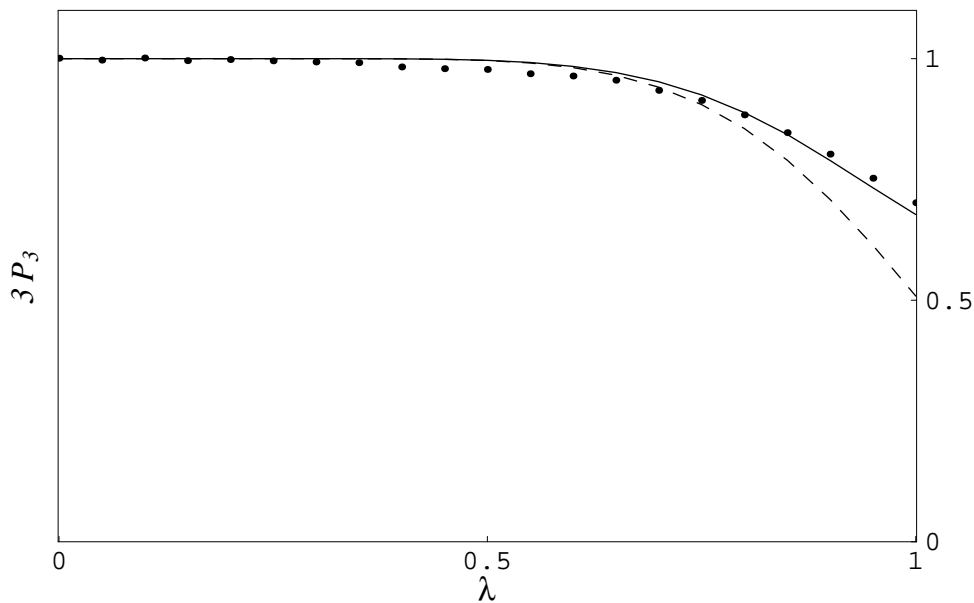}}
\caption{\small Numerical value of $3 P_3$ (dots) and the theoretical prediction using
 equations (\ref{eqPuno}) and 
(\ref{eqPdos}) with the corrected values of $q_1$ and $q_2$ (continuous line) and without the
 corrections (discontinuous line).}
\label{p3teo}
\end{figure}

This procedure, however, fails to predict the numerical
results in two different aspects.
First, if we try to fit $P_3$ we obtain a large deviation
with respect to the numerical value near the symmetric point.
This is shown in Fig. \ref{p3teo} where the dots represent the 
numerical value and the dashed line represents 
the theoretical prediction obtained as outlined above. 
Also the disappearance of even cycles at $\lambda=1$,
as shown in fig. \ref{p41},
is not taken into account within this approximation
i. e. the theoretical value for $P_4$ does not vanish at $\lambda=1$.
These two facts happen to be connected and will be 
discussed in the next paragraph.

Is is first important to understand why even cycles
disappear when $\lambda=1$. 
The reason is very simple, for if we had a
cycle of even length i. e.  $\sigma(i_m)=i_{m+1}, m=1,\dots,2L+1$,
with $i_{m}\not=i_{m'}$ except $i_1=i_{2L+1}$, 
then either the links in odd position
$d_{i_{2l-1}i_{2l}}$ 
or those in even position 
$d_{i_{2l}i_{2l+1}}$ 
have a smaller sum. 
Assume that
$$\sum_{l=1}^L d_{i_{2l}i_{2l+1}}< \sum_{l=1}^L d_{i_{2l-1}i_{2l}},$$
then the new permutation $\sigma'$ which is equal to $\sigma$ except
for $\sigma'(i_{2l+1})=i_{2l},\quad l=1,\dots,L$ gives a smaller
total distance. To see this, it is enough to realise that, 
given that $M_{1}$ is a symmetric matrix, the sum of the odd links 
for $\sigma$ is replaced by that of the even links in $\sigma'$
which lowers the total distance. Hence it is impossible
to have cycles of even length larger than two, in the optimal permutation
of a symmetric {\it distance} matrix.

The mechanism for disappearance of even cycles we
outlined in previous paragraph can be stated by saying that
$2L$-cycles break into $L$ two-cycles. This is the key point
behind the improvement of the approximation  
in order to account for small cycles.
The idea is that in equations (\ref{eqPuno}) 
and (\ref{eqPdos}) instead of using the value of $P_2$
obtained in the numerical simulations we subtract to it
the two-cycles that come from what would be cycles of even length.
The procedure is then clear: we start with a value for $q_1$ and $q_2$,
say $P_1$ and $2P_2$, we compute with this values the theoretical
number of cycles of even length and subtract from it the real one
obtained in the numerical simulations. These are the cycles that break
into a number of two cycles. We subtract this number from $P_2$, 
introduce the new value of $P_2$ into equation (\ref{eqPdos})
and compute again $q_1$ and $q_2$. The procedure is iterated until the
desired convergence is reached. In practise in 4 or 5 iterations we obtain
a very good precision.

\begin{figure}[h!]
\resizebox{130mm}{80mm}{
\includegraphics  {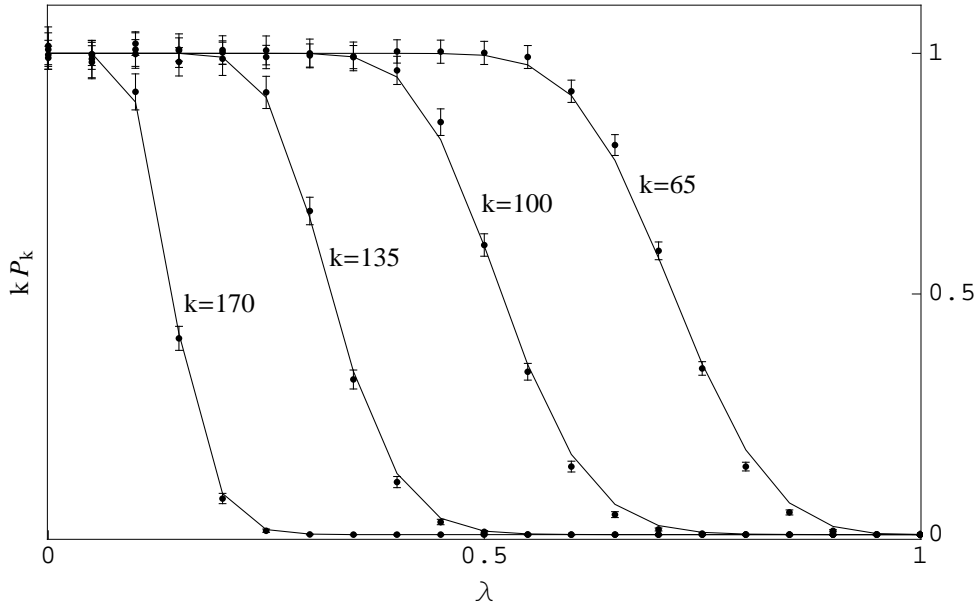}}
\caption{\small Values of $kP_k$ for intermediate values of $k$,
 $N=200$ and $\lambda > 0$. The continuous line is the theoretical prediction. }
\label{kpkteo}
\end{figure}

In Fig. \ref{p3teo} we plot the numerical values for $3P_3$
(dots) and  the theoretical curves using the uncorrected version 
for $q_1, q_2$ (dashed line) 
and the corrected ones (solid line). 
We see that the fit is much better in the second instance. 
The theoretical prediction can be also applied to 
the intermediate cycles as shown in Fig. \ref{kpkteo}.
The theoretical 
and numerical values for $kP_k$ with $N=200$
using the corrected $q_1$ and $q_2$, show a very good 
agreement.

\begin{figure}[h!]
\resizebox{130mm}{80mm}{
\includegraphics  {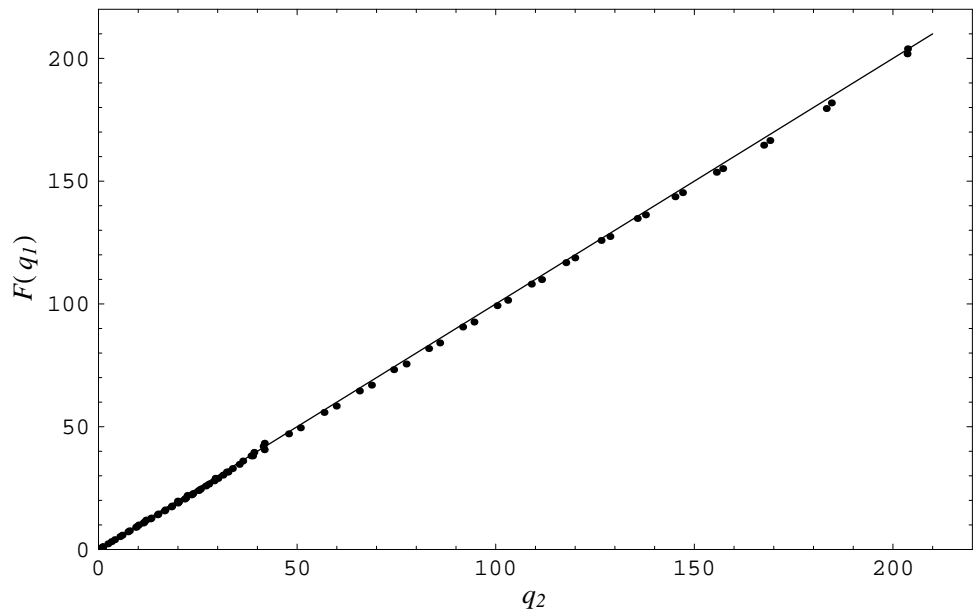}}
\caption{\small Values of $q_2$ versus $F(q_1)=\ee^\lambda q_1(q_1-\lambda)$ for positive  
$\lambda$. The plot includes the points obtained for 
 $N=40,200$ and with the probability density  $\rho = \rho_u$ and $\rho =\rho_e$. }
\label{q1q2+}
\end{figure}

Our last point is the relation between $q_1$ and $q_2$ that 
extends for positive values of $\lambda$ the fit shown in
fig. \ref{q1q2}.
We find that the dependence changes in this case. 
A very good fit is obtained by taking 
$$ q_2=\ee^\lambda q_1(q_1-\lambda)\equiv F(q_1).$$
 As it is shown in fig. \ref{q1q2+} the agreement is rather good and 
it gets better in the large $N$ limit.
 
\section{Conclusions and outlook.}

The expected number of $k$-cycles in the optimal 
permutation of the assignment problem for random matrices,
can be understood to great accuracy in terms of only
two parameters, $q_1$ and $q_2$ associated to one and two-cycles.
More precisely, the ansatz is that in the large $N$ limit
the probability for a permutation to be the solution of the $AP$ 
is proportional to $q_1^{p_1}q_2^{p_2}$, with $p_1, p_2$ the number
of one and two-cycles of the permutation respectively. 
The ansatz can be substantiated by considering that with the cost or 
{\it distance} matrices used in the paper only one and two-cycles
are sensible to the symmetry of the matrix, as bonds of longer cycles are 
uncorrelated. On the other hand in the large $N$ limit we can consider
the occurrence of short cycles as independent events.  

With this ansatz we are able to explain, with great accuracy,
the expected number of $k$-cycles in the solution of the AP for
cost matrices ranging from the symmetric to the antisymmetric one.  
The parameters suffer an abrupt 
transition (in the large $N$ limit)
when moving from a matrix mostly symmetric ($\lambda>0$)
to another one mostly antisymmetric ($\lambda<0$).

We also find some universal scaling relations in the variables
which are valid in the antisymmetric region.
Based in this scaling behaviour we are able to give a 
theoretical prediction for the slope of $q_1$ 
at the critical point, $\lambda=0$.

An open problem is to understand the behaviour of the cycles of even 
length in the symmetric region. It is clear that, as it is argued
in the paper, all of them  (except the two cycles) should be absent 
at the symmetric point 
($\lambda=1$), but for the moment we do not know
how to explain the curves that the average number 
of even cycles follow to reach the zero value.
Finally, it would be nice to have a full theoretical study of the model
(or a reliable approximation to it) that could 
explain the facts mentioned above.

\noindent{\bf Acknowledgements:} Research partially supported by
grants FIS2006-01225 and FPA2006-02315, MEC (Spain).


\end{document}